\documentstyle[aps,prl,twocolumn,epsfig]{revtex}

\begin{document}
\twocolumn[\hsize\textwidth\columnwidth\hsize\csname@twocolumnfalse\endcsname

\title{Hydrodynamic modes in a trapped gas of metastable helium above the
Bose-Einstein transition}
\author{M. Leduc, J. L\'eonard, F. Pereira dos Santos, E. Jahier, S. Schwartz and C. Cohen-Tannoudji\\
{\it Coll\`ege de France et Laboratoire Kastler Brossel,\\ D\'epartement de Physique, \'Ecole Normale Sup\'erieure, \\
24 rue Lhomond, 75231 Paris Cedex 05, France}}
\date{\today}
\maketitle

\begin{abstract}

This article describes experiments performed with an ultracold gas of
metastable helium above the Bose-Einstein transition. The gas is
trapped in an harmonic magnetic potential and collective excitations
are induced. Both the frequency and the damping rate of the lowest
monopole-quadrupole $m=0$ mode of excitation are studied at different
temperatures and compared to theoretical predictions. Results indicate
that the gas goes from a collisionless regime towards a hydrodynamic
regime as the elastic collision rate increases, when one goes down
along the evaporative cooling ramp. However we  find a discrepancy for
the collisional parameters in comparison with predictions relying on
the value of the scattering length previously estimated. Possible
explanations are put forward in the final discussion.

\

\

\end{abstract}]

\section{Introduction}

Soon after the experimental realization of Bose-Einstein condensation
in trapped atomic gases, studies of collective excitations in these
systems have become a subject of intense investigation by several
groups \cite{Giorgini97,JILA97,Stamper98,Dafolvo99,Griffin97,Guery99}.
At very low temperature, when the system is Bose-Einstein condensed, it
can be described by the hydrodynamic equations of a superfluid
\cite{Dafolvo99,Stringari96}, which give predictions in good agreement
with experiments performed with alkali condensates. At higher
temperature, the mean field effects become less important and
collisions turn out to be the dominant factors. For an ultracold gas
above the critical temperature of condensation, the dynamical behavior
of the dilute gas can be described by the Boltzmann equation. In this
case two extreme situations can occur: - the hydrodynamic regime, where
the mean free path  between the colliding particles is small compared
to the dimensions of the cold cloud confined in its trap - and the
collisionless regime where the motion is described by the single
particle Hamiltonian. In both cases, when excited, the system shows
well-defined oscillations resulting from the external magnetic
confinement, for which predictions have been made (see for instance
\cite{Griffin97} or \cite{Guery99} and references therein). The
frequency and the damping rates of the gas oscillations can be
calculated for a classical gas confined in an harmonic trap, neglecting
the mean field effects. In reference \cite{Guery99} results are given
for the monopole-quadrupole modes of excitation, both for the
hydrodynamic and for the collisionless regime, with an explicit
description of the transition between the two regimes. Many
experimental studies have focused on collective excitations of such
gases. Experiments have studied low-lying excitations over large ranges
of temperature.  For all the available data dealing with the thermal
cloud of trapped alkali gases significantly above the critical
temperature $T_{c}$ the regime is collisionless, due to the relatively
low value of the atomic density and of the $s$-wave scattering length
ruling the elastic collision rate at very low temperatures. Experiments
performed at MIT with a dense gas of sodium provided oscillations of
the thermal cloud analogous to first sound and approaching the
hydrodynamic regime \cite{Stamper98}. However this regime was never
completely reached and the complete check of the predictions of
\cite{Guery99} is still to come.

When Bose-Einstein condensation of metastable helium was first achieved
in 2001 \cite{Robert01,Pereira01}, it was soon realized that the large
estimated value of the scattering length (16 $\pm$ 8 nm) would favor a
situation where the hydrodynamic regime could be observed in the
thermal cloud above $T_{c}$. In particular our experiment at ENS seemed
appropriate for such a study because of its relatively large number of
atoms at the transition ($8\times 10^{6}$). We thus decided to focus on
the study of collective excitations in the thermal cloud above $T_{c}$.
The present article first briefly recalls the experimental set-up and
the quantitative results already obtained for the relevant parameters:
atomic density, temperature, elastic collision rate at the end of the
evaporative cooling ramp. Then we explain the generation of excitations
in the cigar shaped thermal cloud and describe the optical measurements
of the oscillations of the monopole-quadrupole $m=0$ mode generated
along the weak axis of the cloud. The data for the damping and the
frequency of the mode are compared with the predictions of reference
\cite{Guery99} and a discussion follows. In particular one finds a
discrepancy between the elastic collision rate estimated from the
initial data of \cite{Pereira01} and those actually measured through
the present method.

\section{Magnetic trapping and optical detection}

The production of a Bose-Einstein condensate of $^{4}$He atoms in the
metastable $2^{3}S_{1}$ state at ENS is first reported in
\cite{Pereira01} and described in more detail in \cite{Pereira02}. The
metastable atoms are produced from a cryogenic discharge source at a
rate of $10^{14}$ at/s/sr. The atomic beam is manipulated by a laser
operating on the $2^{3}S_{1}-2^{3}P_{2}$ transition of the helium atom
at 1.083 $\mu$m. After transverse collimation, deflection and
longitudinal deceleration in a 2 m long Zeeman slower, the beam loads a
magneto-optical trap using laser beams with a large detuning
compensated by a large intensity, in order to minimise the rate of
inelastic Penning collisions induced by the laser
\cite{Bardou92,Mastwijk98,Kumakura99,Tol99,Pereira011,Browaeys00}.

\begin{figure}[htb]
\begin{center}
\epsfig{file=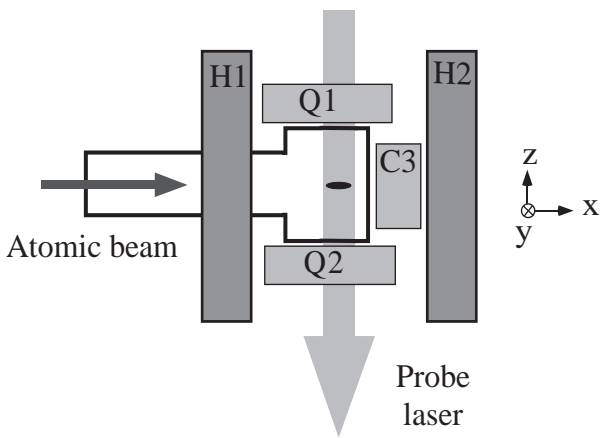,height=5.7cm,width=8cm}
\end{center}
\caption{\footnotesize Magnetostatic trap for the harmonic
confinement of the cold helium gas. Q1, Q2, C3 are the
Ioffe-Pritchard coils, H1 and H2 are the Helmholtz coils for the
bias compression. The detection is optical, using the absorption
of a probe laser beam imaged on a CCD camera.} \label{FigTrap}
\end{figure}

The atomic cloud of typically $10^{9}$ atoms at 1 mK is then
transfered into a magnetic trap through a multistep process
\cite{Pereira02} with an efficiency of about 75\%. The magnetic
trap that we use for the confinement is sketched in
figure~\ref{FigTrap}.

Three coils Q1, Q2, C3 of relatively small size realize an anisotropic
magnetic Ioffe-Pritchard trap. Two additional Helmholtz coils H1 and H2
reduce the bias field in order to increase the radial confinement. A
current of  46.6 A in all five coils produces a 5 G bias field, radial
gradients of 266 G/cm and an axial curvature of 186~G/cm$^2$. The trap
depth is then about 16 mK, the trap frequency along the $x$ axis
($\parallel$) is $\omega_{\parallel}/2\pi=\omega_{x}/2\pi=115$~Hz. The
radial frequency is $\omega_{y}/2\pi= \omega_{z}/2\pi=190$~Hz without
compensation of the bias and 988 Hz with a 5 G bias field. The cold
cloud at thermal equilibrium has a cigar shape, elongated along the $x$
axis and with a cylindrical symmetry in the radial directions along the
$y$ and $z$ axis. When required the cloud can be further compressed in
the radial direction: for this we use another pair of Helmholtz coils
delivering an additional field along the $x$ axis which reduces the
value of the bias field below 5 G down to nearly 0. The current in the
5 coils can be switched off in 200 $\mu$s, but eddy currents induced in
several pieces of the set-up in the vicinity of the cell create
relatively strong transient field gradients which disappear with a time
scale of a few ms. One has to take into account this phenomenon which
can create some difficulties when deducing quantitative values from
time of flight measurements.

The optical detection is shown in Fig. \ref{FigTrap}. One measures the
absorption of a laser probe beam passing through the atomic cloud with
a CCD camera (Hamamatsu C4880) with 1.5\% efficiency at the wavelength
$\lambda=$1.083 $\mu$m. The laser source is a DBR diode laser (SDL
60702 H1) operated in an extended cavity and amplified by an Yb doped
amplifier. The probe pulse is long enough (100 $\mu$s) and intense
enough ($I=0.2 I_{sat}$) to give a reasonable signal to noise ratio in
spite of the low efficiency of the detector. The magnification being of
the order of 1, the resolution of the images is given by the size of
the detector pixel, namely 24 $\mu$m. Usually the image of the cold
cloud in the trap is too small to be detected with a good resolution.
Thus after switching off the trap, we let the cloud expand for a few ms
and take a picture. We deduce from the images both the actual size of
the cloud in the trap and its temperature.

The temperature is ramped down in the magnetic trap by evaporative
cooling. Details about the optimisation of the ramp are given in
reference \cite{Pereira02}. In 8 s one decreases the temperature down
to the $\mu$K range, whereas the phase space density increases by 6
orders of magnitude and one reaches the Bose-Einstein condensation
around 5 $\mu$K. At the threshold of the transition the dimensions of
the cloud have decreased to 90 and  9 $\mu$m in the axial and radial
directions, respectivrly.

\section{Relevant parameters}

\subsection{Number of atoms}

An important parameter for collision studies is the total number of
atoms. As already discussed in \cite{Pereira01}, a proper calibration
of this number $N$ is difficult to extract with a good precision from
the absorption images, due to various difficulties dealing with the
duration of the pulse, the Penning collisions that it induces, its
pushing of the atoms, and stray magnetic fields produced by eddy
currents. We are currently developing a calibration method based on
fluorescence measurements that will be published elsewhere. For the
time being we prefer to rely on the method discussed in
\cite{Pereira01} based on the determination of the critical temperature
$T_{c}$ for the condensation. $T_{c}$ is deduced from measurements of
the relative number of atoms in the condensate and in the thermal cloud
as a function of temperature. The determination of the critical
temperature does not require knowing the absolute number of atoms in
each phase. However it can be used to evaluate the number of atoms
$N_{c}$ at the transition, assuming that one deals with an ideal
bosonic gas. If so, $N_{c}$ and $T_{c}$ are related by:
\begin{equation}
   N_{c}= 1.2 (k_{b}T_{c}/h \overline{\nu})^{3}
\end{equation}
where $h$ is the Planck constant, $k_{b}$ the Boltzmann constant and
$\overline{\nu}$ the geometrical average of frequencies of the trap
(482 Hz). Measurements of reference \cite{Pereira01} gave $T_{c} = 4.7
\pm 0.5\, \mu$K. One thus deduces $N_{c} = 8.2\times 10^{6}$ atoms with
an uncertainty of 30\%. We neglected the possible errors resulting
either from mean field interactions or from quantum correlations (see
the discussion of note 13 in reference \cite{Pereira01}). Once
calibrated at $T_{c}$ the number of atoms $N$ is deduced for any
temperature above $T_{c}$.

\subsection{Scattering length}

The scattering length $a$ is another important parameter relevant for
the present studies, as the scattering processes for metastable helium
atoms occur only in the $s$-wave channel at very low temperatures.
There are several ways of estimating the value of $a$ from the present
experiment.  All of them suppose that the absolute number of atoms is
known. As the uncertainty on this number is large, the value of $a$ is
also derived with a large error bar. We here chose to rely on the
method described in \cite{Pereira01}, in which the scattering length is
deduced from the size of the condensate. Its value is determined from
the characteristic size of the ground state of the trap and from the
chemical potential $\mu$, which can be calculated from the size of the
condensate in the Thomas-Fermi limit \cite{Dafolvo99}. We then find the
value $a = 16 \pm 8$ nm, consistent with the experimental measurement
of \cite{Robert01} and theoretical calculations of \cite{Fedichev96}
and \cite{Venturi00} estimating $a$ of the order of 8 nm (with no error
estimate). A recent calculation \cite{Gadea02} is also compatible with
the present experimental value but does not give more indication on its
precision.

\subsection{Elastic collision rates}

Finally from the previous values of parameters $N_{c}$ and $a$,
one can deduce a value for the elastic collision rate
$\Gamma_{coll}$ in the thermal cloud just at the transition before
condensation. We use the simple formula:
\begin{equation}
    \Gamma_{coll}  =   \overline{n}\sigma \overline{v}
\end{equation}
where $\overline{n}$ is the average density and $\overline{v}$ is
the mean relative velocity of the atoms at the transition
temperature of 4.7 $\mu$K. One easily finds that:
$$             \overline{v} = 4\sqrt{\frac{k_{b}T}{\pi m}}= 23 \: \, {\textrm{cm\,s}}^{-1}$$
and $$\overline{n} = 1.4\times 10^{13}{\textrm{cm}}^{-3}.$$ The value
of the cross section $\sigma$ is estimated from that of the scattering
length by:
$$        \sigma = 8 \pi a^{2} =  6.4 \times 10^{-11}  {\textrm{\, cm}}^{2}$$ which finally
results in:
\begin{equation}
\Gamma_{coll}=2 \times 10^{4} {\textrm{s}}^{-1} .
\end{equation}
The uncertainty on this value is estimated within a factor of 4 which
combines those on $a$, on $N_{c}$ and on $T$. However it can be
compared to the low angular frequency of the trap ($\omega_{\parallel}=
2\pi\times 115$~Hz), and one finds a ratio $ \omega_{\parallel} /
\Gamma_{coll} = 0.04$, much smaller than 1. A collision rate higher
than the trap frequency also means that the mean free path
$L\sim1/\overline{n}\sigma$ of the atoms between collisions is smaller
than the dimensions of the cloud. Near the critical temperature, $L
\sim 11\,\mu{\textrm m}$, which is smaller than the size of the cloud
along the weak axis $\sigma_{\parallel}\sim 140 \,\mu$m. We can thus
assume that one enters into the hydrodynamic regime, an interesting
feature for a cold gas above the transition, as stated in the
introduction. These results motivated us to undertake the following
studies of the collective excitations in the thermal cloud above
$T_{c}$.

\section{Generation and detection of excitation modes}

We decided to generate the lowest excitation modes in the cold gas. The
goal of this study is to observe the oscillations of the cold cloud in
the axial direction after it receives a kick appropriate to generate
the monopole-quadrupole $m=0$ mode.  Let us recall that in an
anisotropic trap the modes $l=m=0$ (monopole) and $l=2$, $m=0$
(quadrupole) are coupled and this results in two so-called '{\it
monopole-quadrupole}' modes. We study the low-frequency
monopole-quadrupole mode, for which the axial and radial oscillations
are in opposite phase.

The calculations of reference \cite{Guery99} predict changes for the
frequency $\omega_{Q}$ and for the damping $\Gamma_{Q}$ of the
oscillation mode when the collision rate varies from a collisionless
regime to a hydrodynamic regime (see Fig. \ref{Figtheory}). We thus
decided to measure the response in time of the ellipticity of the cloud
after the mode is excited. We repeated the measurements at various
temperatures above $T_{c}$ in a large range of collision rates,
expecting to reach the hydrodynamic regime at the vicinity of the
condensation transition.

The excitation is made by a transient pulse of magnetic field giving a
kick to the trapped cloud. Several procedures were used and finally we
found that the most efficient one is the following one:

- One starts with the gas at rest in the trap at a given temperature
$T$ above $T_{c}$ in the conditions used for the evaporative cooling
ramp, namely with a bias field $B_{0}$ of 5 G resulting of an equal
current in the 5 coils of the trap.

- One abruptly lowers the bias field $B_{0}$ along the weak axis ($x$
direction) to a fraction of a Gauss by applying an appropriate current
to the additional bias coils mentioned above. This creates a
compression in the radial direction, the cloud becomes very elongated.
The trap is no longer harmonic but the atomic density increases
dramatically.

- One simultaneously creates a sinusoidal modulation of this
compression field with the additional bias coils. The amplitude of the
short modulation pulse is of  order 0.15 G, its duration of order 15 ms
(3 periods). We chose a modulation frequency of 180 Hz, intermediate
between 2 $\nu_{\parallel}$ and 1.55 $\nu_{\parallel}$ which are the
expected frequencies for the low frequency monopole-quadrupole mode in
the two extreme cases, the collisionless regime and the hydrodynamic
regime \cite{Guery99}.

- Just after this pulse one changes the bias field again and sets it to
about 2 G, a value significantly lower than the usual bias field
$B_{0}$ of 5 G. This value is chosen in order the obtain the maximum
compression of the gas compatible with an harmonic trap. On the one
hand the higher the compression, the larger the collision rate,
facilitating the entrance into the hydrodynamic regime. On the other
the trap has to remain harmonic, meaning that the condition $\mu
B/k_{b}T \gg 1$ must remain fulfilled, to avoid that the intrinsic
frequency of the trap is changed by the anharmonicity.

One can show that such a procedure allows to excite the low frequency
monopole-quadrupole mode $m=0$.

\section{Experimental results, comparison with theory}

\begin{figure}[bht]
\begin{center}
\epsfig{file=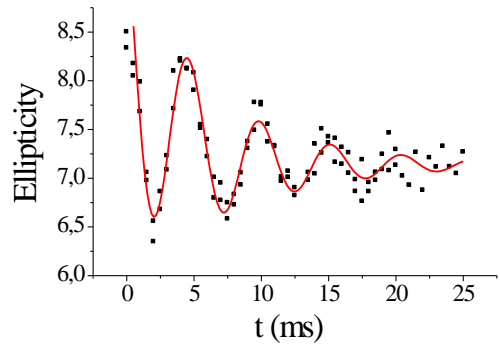,height=4.5cm,width=6.5cm}
\end{center}
\caption{\footnotesize Oscillations of the ellipticity of the atomic
cloud in the monopole-quadrupole $m=0$ mode. The frequency and damping
given by the fit are $\omega_{Q}/\omega_{\parallel} =1.66$ and
$\Gamma_{Q}/\omega_{\parallel}=0.22$.}  \label{FigOscill}
\end{figure}

We monitor the evolution of the ellipticity of the cloud as a function
of time after the oscillation is produced. Actually the ellipticity
oscillates at the same frequency as the collective mode with the same
damping in case of a sufficiently small excitation. A measurement is
shown in Fig.\ref{FigOscill}.

From such data one extracts the oscillation frequency $\omega_{Q}$ and
the damping rate $\Gamma_{Q}$ of the excitation mode. For instance the
results of Fig. \ref{FigOscill}, taken at $3T_{c}$ , give
$\omega_{Q}/2\pi =191 \pm 6$Hz and $\Gamma_{Q}/2\pi =25.3\pm 0.8$ Hz.
This experiment is repeated several times at different values of the
elastic collision rate obtained by varying the final frequency of the
RF knife of the evaporation ramp.

\begin{figure}[htb]
\begin{center}
\epsfig{file=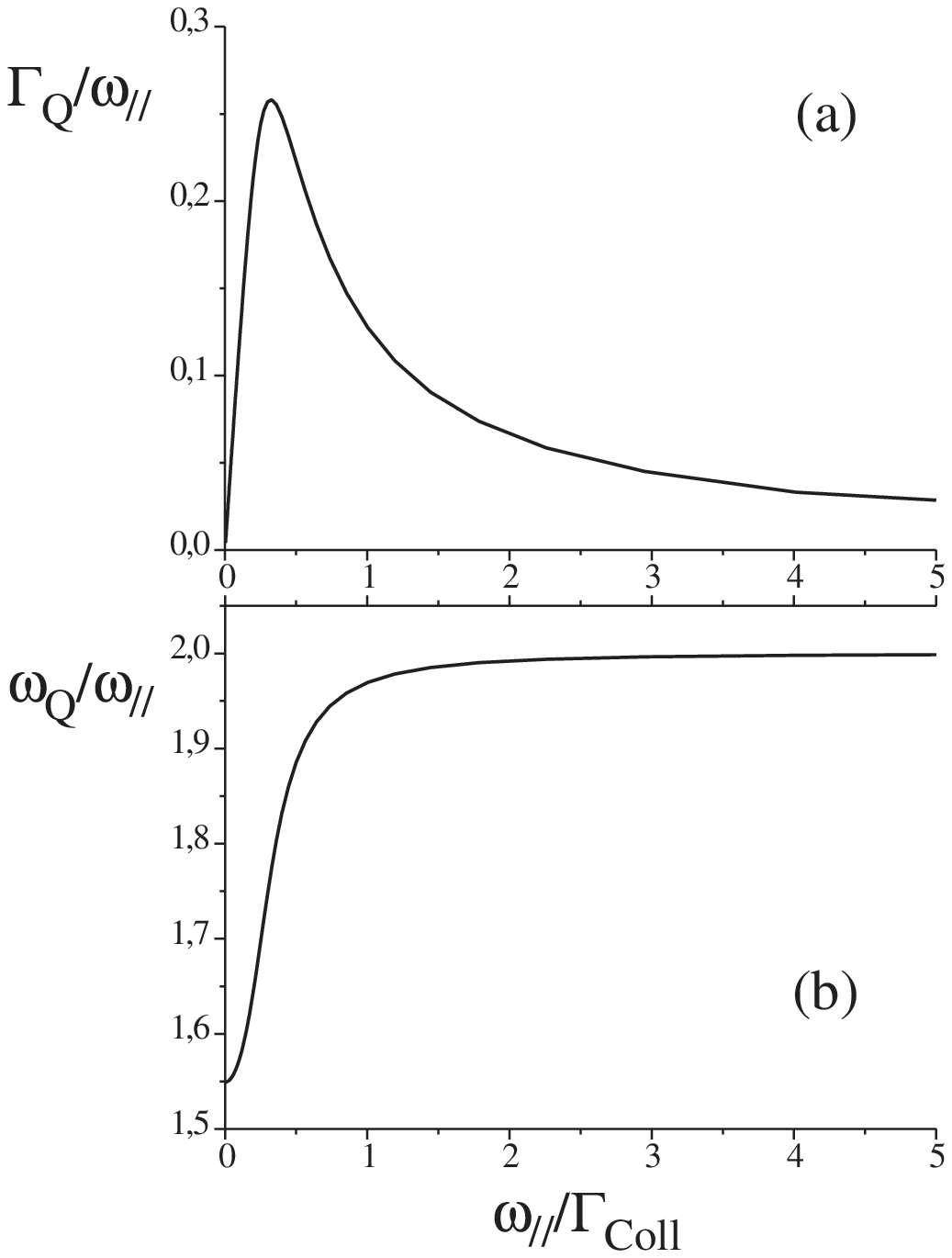,height=9cm,width=6.5cm}
\end{center}
\caption{\footnotesize Theoretical predictions of [6] for the low
frequency monopole-quadrupole mode $m=0$ of an elongated cloud of cold
gas as a function of the elastic collision rate. On the upper part
(Fig. 3a) is plotted the damping of the oscillations, and on the lower
part (Fig.3b) the frequency of the oscillations, both in units of
$\omega_{\parallel}$, the axial (weak) frequency of the trap (Figure
taken from [6]).}\label{Figtheory}
\end{figure}

The results are compared with the predictions of the classical
theory of reference \cite{Guery99}, which are recalled in Fig.
\ref{Figtheory}.

\begin{figure}[htb]
\begin{center}
\epsfig{file=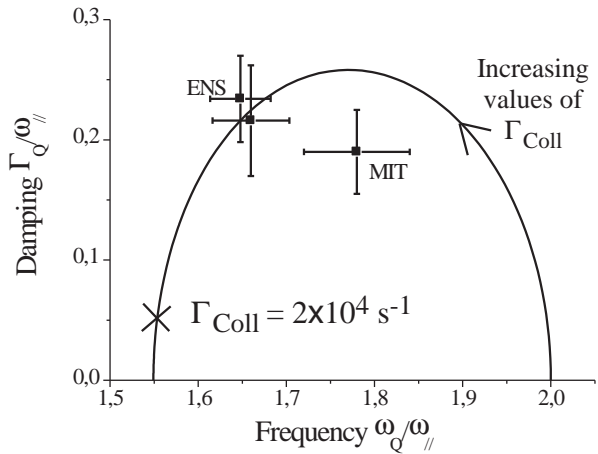,height=6cm}
\end{center}
\caption{\footnotesize Plot of the frequencies and damping rates of the
lowest monopole-quadupole mode m=0. The solid line summarizes the
theoretical calculation already mentioned in Fig.\ref{Figtheory}. The
cross corresponds to the estimated value of the collision rate
$\Gamma_{coll} = 2 \times 10^{4} s^{-1}$ for metastable helium as
derived from [9]. The arrow indicates the displacement on the curve
when $\Gamma_{coll}$ increases. ENS: Experimental results of this work
for metastable helium. MIT: Experimental data point of [3] for sodium.
}
 \label{figResult}
\end{figure}

These results derive both from numerical simulations and from
analytical calculations, which are in excellent agreement. One notes
that the ratio $\omega_{Q} /\omega_{\parallel}$ decreases from 2 for
the collisionless regime to 1.55 for the hydrodynamic regime, whereas
$\Gamma_{Q}/\omega_{\parallel}$ goes to 0 at {\it both} ends but
reaches a maximum in between the two regimes. The results of Fig.3a and
Fig.3b can be combined to give the curve of Fig.\ref{figResult}.

Fig.\ref{figResult} shows the experimental data we obtained. The
advantage of such a presentation is that only relative values of the
measured frequencies are used, which are known with good precision. One
does not require knowledge of the collision rates, which is difficult
due to the uncertainties in the present stage of the experiment. In
Fig. \ref{figResult} we plotted the points (called "ENS") corresponding
to the highest collision rates that we were able to achieve at a
temperature of about $3T_{c}$. The comparison between experimental data
and theoretical calculation allows us to make an indirect measurement
of the elastic collision rate in the excited cloud. The highest
collision rate that we obtained is in the range of $3.5 \times 10^{3}$
s$^{-1}$. One remarks that we were unable to reach the expected
collision rate of $2 \times 10^{4} s^{-1}$ which is indicated with a
cross in Fig. \ref{figResult} and which was estimated independently at
the transition \cite{Pereira01}. We also plotted the results obtained
by the group of W. Ketterle at MIT
 \cite{Stamper98} for cold Na atoms.\\

\section{Discussion}

\begin{figure}[htb]
\begin{center}
\epsfig{file=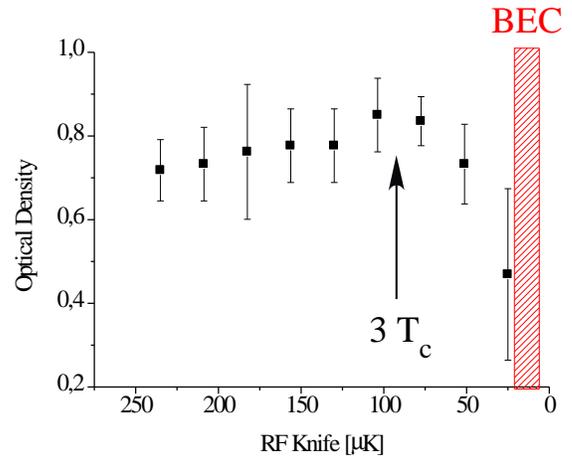,height=6 cm}
\end{center}
\caption{\footnotesize Measurement of the optical density at the center
of the cloud as a function of the trap depth at the end of the
evaporative cooling ramp. The optical density being proportional to the
elastic collision rate in a harmonic trap, this plot shows that one
leaves the run-away regime below a temperature of about $3 T_{c}$.}
\label{figevap}
\end{figure}

Usually, when the initial elastic collision rate is high enough, the RF
evaporation ramp enters in the so-called run-away regime, where both
phase space density and $\Gamma_{coll}$ increase exponentially. So we
expect to measure the highest collision rate at the Bose-Einstein
transition. In our case, we observed that the phase space density
always increases (leading to the Bose-Einstein condensation) but on the
contrary, at the very end of the evaporation ramp, the collision rate
starts to decrease. The two behaviors are not incompatible since the
phase space density scales like $N/T^{3}$ whereas the collision rate
scales like $N/T$. Monitoring the evolution of the elastic collision
rate along the evaporation ramp can be easily done while monitoring the
optical density ($OD$) at the center of the cloud. Actually, both
quantities are proportional in a harmonic trap : $OD \propto N/T
\propto \Gamma_{Coll}$. The evolution of the optical density of the
cloud during the evaporative cooling is plotted in Fig. \ref{figevap}.
This plot shows that the highest elastic collision rate we can produce
is obtained at about $3T_{C}$. Then, instead of getting deeper into the
hydrodynamic regime, the elastic collision rate turns back towards
lower values while approaching the transition.

Actually, while studying the monopole-quadrupole excitation, the
highest elastic collision rate measured in Fig. \ref{figResult}
($\Gamma_{Coll}^{max} \sim 3.5 \times 10^{3}$ s$^{-1}$) was obtained
after evaporation down to about $3T_{C}$, and {\it after additional
compression} of the magnetic trap. From this value we can infer the
corresponding elastic collision rate before additional compression : $
2 \times 10 ^{3}$ s$^{-1}$. Then, from Fig. \ref{figevap} we can infer
the collision rate at the critical temperature $T_{C}$ which is about
twice as small, leading to $10^3$ collisions per second. This value is
about 20 times smaller than the one derived from the measurement of the
critical temperature of the phase transition.

Within our error bars, assuming 8 nm for the scattering length instead
of 16 nm reduces by 4 the discrepancy between the two independent
measurements of the elastic collision rate. Then, when we first
estimated the collision rate knowing the critical temperature, the
number of atoms at the transition was nearly twice as large as it was
when we produced collective excitations, which also explains part of
the discrepancy. Still it is not enough to conclude to the agreement
between the two values. We are still working on improving the accuracy
of both kinds of measurement to understand the discrepancy. In
particular we started studying in more detail the effects of eddy
currents on the measurement of the temperature. In fact, magnetic
fields remaining after the switching off of the trap Zeeman shift the
optical lines and modify the absorption cross section of atoms during
the optical detection. This modification is not homogeneous in space
and might lead to a systematic error on the effective measurement of
the size of the cloud after free expansion, and, consequently, of the
critical temperature.

In order to find out why we leave the run-away regime at the end of the
evaporative ramp, we studied losses and heating rates. For instance we
measured heating rates of 20 $\mu$K/s at $3T_{c}$. A possible
explanation is related to inelastic Penning Ionization taking place
when the density becomes high enough. A dominant process is the
formation of He$^{+}$ ions and He atoms in the ground state which have
very large kinetic energy originated from the internal energy of the
metastable atoms (about 20 eV). These hot products might collide with
trapped metastable atoms and heat them up {\it via} elastic collisions
\cite{Beijerinck01}. Estimates from numbers given in
\cite{Beijerinck01} lead to the right order of magnitude for the
heating rate. The more hydrodynamic, the more efficient the heating
process will be \cite{Rempe01}, apparently leading to a limitation in
the highest elastic collision rate we can produce.

In conclusion, the route towards the hydrodynamic limit has been
explored further than in any other reported experiment on collective
excitation modes. Especially the measured frequency of the excited mode
is strongly shifted towards the hydrodynamic regime as predicted. The
experiment is still under progress, and we hope to go deeper into the
hydrodynamic regime. However, intrinsic limitations due to inelastic
collisions have been found.

\

\noindent {\bf Acknowledgements:} The authors wish to thank the writers
of \cite{Guery99} for allowing the reproduction of part of their work
(Fig. \ref{Figtheory}). This work was supported by La R\'egion
Ile-de-France through SESAME contract number 521027.


\begin{thebibliography}{30}
\bibitem{Giorgini97} S. Giorgini, L. P. Pitaevskii, S. Stringari, {\it Phys. Rev. Lett.} {\bf 78}, 3987 (1997)

\bibitem{JILA97} D. S. Jin, M. R. Matthews, J. R. Ensher, C. E. Wieman, and E. A.
Cornell, {\it Phys. Rev. Lett.} {\bf 78}, 764 (1997)

\bibitem{Stamper98} D. M. Stamper-Kurn, H.-J. Miesner, S. Inouye, M. R. Andrews, W. Ketterle,
{\it Phys. Rev. Lett.} {\bf 81}, 500 (1998)

\bibitem{Dafolvo99} F. Dafolvo, S. Giorgini, L. P. Pitaevskii, S. Stringari, {\it Rev. Mod. Phys.} {\bf 71}, 463 (1999)

\bibitem{Griffin97} A. Griffin, Wen-Chin Wu, S. Stringari, {\it Phys. Rev. Lett.} {\bf 78}, 1838 (1997)

\bibitem{Guery99} D. Gu\'ery-Odelin, F. Zambelli, J. Dalibard, S. Stringari, {\it Phys. Rev. A} {\bf 60}, 4851 (1999)

\bibitem{Stringari96} S. Stringari, {\it Phys. Rev. Lett.} {\bf 77}, 2360 (1996)

\bibitem{Robert01} A. Robert, O. Sirjean, A. Browaeys, J. Poupard, S. Nowak, D. Boiron, C.I. Westbrook, A. Aspect,
{\it Science Mag.} 292, 463 (2001)

\bibitem{Pereira01} F. Pereira dos Santos, J. L\'eonard, Junmin Wang, C. J. Barrelet, F. Perales,
E. Rasel, C. S. Unnikrishnan, M. Leduc, C. Cohen-Tannoudji, {\it
Phys. Rev. Lett.} {\bf 86}, 3459 (2001)


\bibitem{Pereira02} F. Pereira dos Santos, J. L\'eonard, Junmin Wang, C. J. Barrelet, F. Perales,
E. Rasel, C. S. Unnikrishnan, M. Leduc and C. Cohen-Tannoudji, {\it
Eur. Phys. J. D}, {\bf19}, 103 (2002)


\bibitem{Bardou92} F. Bardou, O. Emile, J.M. Courty, C.I. Westbrook and A. Aspect, Europhys. Lett. {\bf 20}, 681 (1992)

\bibitem{Mastwijk98} H.C. Mastwijk, J.W. Thomsen, P. van der Straten and A. Niehaus, Phys. Rev. Lett. {\bf 80}, 5516 (1998).

\bibitem{Kumakura99} M. Kumakura and N. Morita, Phys. Rev. Lett. {\bf 82}, 2848 (1999).

\bibitem{Tol99} P.J.J. Tol, N. Herschbach, E.A. Hessels, W. Hogervorst, W. Vassen, Phys. Rev. A {\bf 60}, R761 (1999)

\bibitem{Pereira011} F. Pereira Dos Santos, F. Perales, J. L\'eonard, A. Sinatra, Junmin Wang, F. S. Pavone, E. Rasel, C. S. Unnikrishnan and M. Leduc,  Eur. Phys. J. D {\bf 14}, 15 (2001)

\bibitem{Browaeys00} A. Browaeys, J. Poupard, A. Robert, S. Nowak, W. Rooijakkers, E. Arimondo, L. Marcassa, D. Boiron, C.I. Westbrook and A. Aspect, Eur. Phys. J. D {\bf 8}, 199 (2000)


\bibitem{Castin96} Y. Castin, R. Dum, {\it Phys. Rev. Lett.} {\bf 77}, 5315 (1996)

\bibitem{Fedichev96} P. O. Fedichev, M. W. Reynolds, U. M. Rahmanov, G. V. Shlyapnikov,
{\it Phys. Rev. A} {\bf 53}, 1447 (1996)

\bibitem{Venturi00} V. Venturi, I. B. Whittingham, {\it Phys. Rev. A} {\bf 61}, 060703 (2000)

\bibitem{Gadea02} F. X. Gad\'ea, T. Leininger, A. S. Dickinson , to be published in {\it J. Chem. Phys.}

\bibitem{Beijerinck01} H. C. W. Beijerinck, E. J. D. Vredenbregt, R. J. W. Stas, M. R. Doery, and J.
G. C. Tempelaars, {\it Phys. Rev. A} {\bf 61} 023607 (2000)

\bibitem{Rempe01} J. Schuster, A. Marte, S. Amtage, B. Sang, G. Rempe,H. C.W.
Beijerinck, {\it Phys. Rev. Lett.} {\bf 87}, 170404 (2001)

\end{thebibliography}
\end{document}